\documentstyle[preprint,prb,aps,epsf,eqsecnum]{revtex}

\newcommand{\be}{\begin{equation}}
\newcommand{\ee}{\end{equation}}
\newcommand{\bea}{\begin{eqnarray}}
\newcommand{\eea}{\end{eqnarray}}
\newcommand{\non}{\nonumber}
\newcommand{\ra}{\rangle}

\begin{document}
\draft
\tightenlines

\title{XXZ Bethe states as highest weight vectors 
of the $sl_2$ loop algebra at roots of unity
}

\author{Tetsuo Deguchi
\footnote{e-mail deguchi@phys.ocha.ac.jp}}
\address{ Department of Physics, Ochanomizu University, 2-1-1
Ohtsuka, Bunkyo-ku,Tokyo 112-8610, Japan}

\date{\today}

\maketitle

\begin{abstract} 
 We prove some part of the conjecture 
  that regular Bethe ansatz eigenvectors 
of the XXZ spin chain at roots of unity are  
highest weight vectors of the $sl_2$ loop algebra.   
Here  $q$ is related 
to the XXZ anisotropic coupling $\Delta$ 
by  $\Delta=(q+q^{-1})/2$, and it is given by a  
root of unity, $q^{2N}=1$, for a positive integer $N$.  
 We show that  regular XXZ Bethe states are annihilated 
  by the generators ${\bar x}_k^{+}$'s, for  any $N$.  
   We discuss, for some particular cases of $N=2$,  
   that regular XXZ Bethe states are 
   eigenvectors of the generators of the Cartan subalgebra, ${\bar h}_k$'s.  
Here the loop algebra $U(L(sl_2))$ is generated by  
${\bar x}_k^{\pm}$ and ${\bar h}_k$ for $k \in {\bf Z}$,  which 
 are  the classical analogues of the Drinfeld generators   
of the quantum loop algebra $U_q(L(sl_2))$. 
A representation  of $U(L(sl_2))$ is called highest weight 
if it is generated by a vector $\Omega$ which is annihilated 
by the generators ${\bar x}_k^{+}$'s  and such that 
$\Omega$ is an eigenvector of the ${\bar h}_k$'s. 
We also discuss the classical analogue of 
the Drinfeld polynomial which  characterizes  the irreducible 
finite-dimensional highest weight representation of $U(L(sl_2))$.    

%
%
%
%
\end{abstract}

\newpage 

\section{Introduction}

The XXZ spin chain is one of the most important exactly solvable 
quantum systems \cite{Bethe,Yang-Yang}. 
The Hamiltonian under the periodic boundary conditions is given by 
\be 
H_{XXZ} =  - J \sum_{j=1}^{L} \left(\sigma_j^X \sigma_{j+1}^X +
 \sigma_j^Y \sigma_{j+1}^Y + \Delta \sigma_j^Z \sigma_{j+1}^Z  \right) \, . 
\label{hxxz}
\ee
Here the parameter $\Delta$ expresses the anisotropic coupling 
between adjacent spins, and it is related to the $q$ parameter of the 
quantum group $U_{q}(sl_2)$ by the relation  $\Delta= (q+q^{-1})/2$.  

\par 
 Recently it was shown  that when $q$ is a root of unity,  
 the Hamiltonian commutes with the generators of 
the $sl_2$ loop algebra \cite{DFM}. The symmetry of the XXZ Hamiltonian 
is enhanced at the root of unity.   The spectra of the XXZ spin chain 
have been studied numerically \cite{FM,Odyssey},
 and some generalizations of the $sl_2$ loop algebra symmetry 
 have also been derived for some trigonometric 
vertex models \cite{Korff}. Furthermore, the spectral degeneracy associated 
with the $sl_2$ loop algebra has been shown for the eight-vertex model 
and some elliptic IRF models \cite{Missing,CSOS}. 
However,  it has been a conjecture  
that the Bethe ansatz eigenvectors are of highest weight 
with respect to the $sl_2$ loop algebra.

 \par 
In this paper we show several rigorous results supporting    
the highest-weight conjecture.   
Based on the algebraic Bethe ansatz method, 
 we show some part of the conjecture  
  that a regular Bethe ansatz eigenvector of the 
XXZ spin chain  should give a highest weight vector 
of the $sl_2$ loop algebra in the sense of the Drinfeld realization of 
the quantum loop algebra $U_q(L(sl_2))$. 
 Here the regular Bethe ansatz eigenvector is defined by 
 such an eigenvector that 
 is constructed from the Bethe ansatz wave-function 
where all the rapidities are finite.

\par 
Let us consider the classical analogue of the Drinfeld 
realization of $U_q(L(sl_2))$ \cite{Chari-P1,Chari-P2}. 
The $sl_2$ loop algebra $U(L(sl_2))$ is generated by the 
classical analogues of the Drinfeld generators,  
${\bar x}_k^{\pm}$ and ${\bar h}_k$ for $k \in {\bf Z}$. 
In the sense of the Drinfeld realization, 
a representation  
of the loop algebra $U(L(sl_2))$ is highest weight 
if it is generated by a vector $\Omega$ which is annihilated 
by the generator ${\bar x}_k$ for any integer $k$ and such that 
$\Omega$ is a common eigenvector of  generators 
of the Cartan subalgebra, ${\bar h}_k$ 's.   
 For the case of any $N$, 
 we show that a regular XXZ Bethe state $|R \ra$ at the root of unity 
 is annihilated  
 by the generators ${\bar x}_k^{+}$'s: ${\bar x}_k^{+} |R \ra = 0$,  
 for all $k \in {\bf Z}$. 
  For some particular case of $N=2$,   
   we show that the regular XXZ Bethe state $|R \ra$ is a common 
     eigenvector of the  generators ${\bar h}_k$'s of the 
     Cartan subalgebra. 
Thus, the proof of the highest-weight conjecture is partially completed 
for the case of $N=2$.  For the vacuum state, we introduce  a method 
for calculating the eigenvalues of the Cartan generators ${\bar h}_k$'s.  
Some version of the highest-weight conjecture 
has been addressed  in Ref. \cite{DFM} from a different viewpoint.   
It is also addressed in Ref. \cite{Odyssey}. 
%
%

\par 
%
We also discuss a method for calculating the classical analogue  
 of the Drinfeld polynomial. Through some examples of the Drinfeld 
polynomial, we suggest a conjecture that  
the irreducible representation of $U(L(sl_2))$ 
generated by a regular XXZ Bethe state  
with $R$ down spins on $L$ sites  
should be  given by $2^{(L-2R)/N}$, 
when $L$ and $R$ are integral multiples of $N$. 
The result is consistent with the 
analytic computation of the degeneracy 
shown for the eight-vertex model and 
some elliptic SOS models associated with 
the eight-vertex model \cite{Missing,CSOS}. 
In association with the $sl_2$ loop algebra,  
the Drinfeld polynomial is also addressed 
in Ref. \cite{Odyssey} with a different method.

\section{$sl_2$ loop algebra and the Drinfeld realization}

\subsection{Loop algebra symmetry of the XXZ spin chain}
\par 
We review some results on the $sl_2$ loop algebra symmetry of the 
XXZ spin chain. We first 
introduce the quantum group $U_q(sl_2)$. 
The generators $S^{\pm}$ and $S^Z$ satisfy the following relations 
\be 
[S^{+}, S^{-}] = {\frac {q^{2 S^Z} - q^{-2 S^Z}} {q-q^{-1}} } \, , \quad  
[S^Z, S^{\pm}] =  \pm  S^{\pm} 
\ee
Here the parameter $q$ is generic. 
The comultiplication is given by 
\be 
\Delta( S^{\pm}) = S^{\pm} \otimes q^{- S^Z} + q^{S^Z} \otimes S^{\pm} \, , 
\quad \Delta(S^Z) = S^Z \otimes I + I \otimes S^Z 
\ee

\par 
We consider the $L$th tensor product 
of spin 1/2 representation $V^{\otimes L}$.  
The representations of the generators $S^{\pm}$ and $S^{Z}$ are given by 
\bea 
q^{S^Z} &= & q^{\sigma^Z/2} \otimes \cdots \otimes q^{\sigma^Z/2}  \non \\
S^{\pm} &= & \sum_{j=1}^{L} 
q^{\sigma^Z/2} \otimes \cdots \otimes q^{\sigma^Z/2} \otimes 
\sigma_j^{\pm} \otimes q^{-\sigma^Z/2} \otimes \cdots \otimes q^{-\sigma^Z/2}   
\eea
Considering the automorphism of the $U_q(L(sl_2))$, 
we may introduce the following operator 
\be 
T^{\pm} = \sum_{j=1}^{L} 
q^{-\sigma^Z/2} \otimes \cdots \otimes q^{-\sigma^Z/2} \otimes 
\sigma_j^{\pm} \otimes q^{\sigma^Z/2} \otimes \cdots \otimes q^{\sigma^Z/2}   
\ee

\par 
Let us now define the operators $S^{\pm(N)}$ and $T^{\pm(N)}$ by   
\be 
S^{\pm(N)} = \left( S^{\pm} \right)^N/[N]! \, , \quad 
T^{\pm(N)} = \left( T^{\pm} \right)^N/[N]!  
\ee
Here we have used the $q$-integer $[n]=(q^n -q^{-n})/(q-q^{-1})$ 
and the $q$-factorial $[n]! = [n] [n-1] \cdots [1]$. 
We also note that $q$ is not a root of unity. 
Then, we can show the following expressions 
\begin{eqnarray}
S^{\pm(N)}&=&  
\sum_{1 \le j_1 < \cdots < j_N \le L}
q^{{N \over 2 } \sigma^Z} \otimes \cdots \otimes q^{{N \over 2} \sigma^Z}
\otimes \sigma_{j_1}^{\pm} \otimes
q^{{(N-2) \over 2} \sigma^Z} \otimes  \cdots \otimes q^{{(N-2) \over 2}
\sigma^Z}
\nonumber \\
 & & \otimes \sigma_{j_2}^{\pm} \otimes q^{{(N-4) \over 2} \sigma^Z} \otimes
\cdots
\otimes \sigma^{\pm}_{j_N} \otimes q^{-{N \over 2} \sigma^Z} \otimes \cdots
\otimes q^{-{N \over 2} \sigma^Z} \quad . 
\label{sn}
\end{eqnarray}
 For $T^{\pm(N)}$ we have 
\begin{eqnarray}
T^{\pm(N)} & =&  
\sum_{1 \le j_1 < \cdots < j_N \le L}
q^{- {N \over 2 } \sigma^Z} \otimes \cdots \otimes q^{- {N \over 2} \sigma^Z}
\otimes \sigma_{j_1}^{\pm} \otimes
q^{- {(N-2) \over 2} \sigma^Z} \otimes  \cdots \otimes q^{ - {(N-2) \over 2}
\sigma^Z}
\nonumber \\
 & & \otimes \sigma_{j_2}^{\pm} \otimes q^{- {(N-4) \over 2} \sigma^Z} \otimes
\cdots
\otimes \sigma^{\pm}_{j_N} \otimes q^{{N \over 2} \sigma^Z} \otimes \cdots
\otimes q^{{N \over 2} \sigma^Z} \quad . 
\label{tn}
\end{eqnarray}
Here we recall that  $q$ is generic. 

\par 
Let the symbol $\tau_{6V}(v)$ denotes the (inhomogeneous) transfer matrix 
of the six-vertex model.  
We now take the parameter $q$ a root of unity.  
We consider the limit of sending $q$ to a root of unity:  $q^{2N}=1$.     
Then we can show the (anti) commutation relations \cite{DFM} 
in the sector of $S^Z \equiv 0~ (\rm{mod}~N)$ 
\begin{equation}
S^{\pm (N)} \tau_{6V}(v)=q^N \tau_{6V}(v) S^{\pm (N)}, \qquad 
T^{\pm (N)} \tau_{6V}(v)=q^N \tau_{6V}(v) T^{\pm (N)} 
\end{equation}
Here we recall that $S^Z$ denotes the $Z$-component 
of the total spin operator. 

\par 
Since the XXZ Hamiltonian $H_{XXZ}$ is given by 
the logarithmic derivative of the (homogeneous) transfer matrix 
$T_{6V}(v)$,    
we have  in the sector $S^Z \equiv 0$ (mod $N$)  
\begin{eqnarray}
{[}S^{\pm(N)},H_{XXZ} {]}={[}T^{\pm(N)},H_{XXZ} {]}=0.
\label{sthcomm}
\end{eqnarray}
Thus, the operators $S^{\pm (N)}$  and $T^{\pm (N)}$ commute 
with the XXZ Hamiltonian in the sector  $S^Z \equiv 0$ (mod $N$) .

\par 
Let us  now consider the algebra generated by 
 the operators $S^{\pm (N)}$ and $T^{\pm (N)}$  \cite{DFM}. 
When $q$ is a primitive $2N$th root of unity, 
or a primitive $N$th root of unity with $N$ odd, 
we consider the  identification in the following \cite{DFM}:    
\begin{equation}
E_0^{+}=S^{+(N)}, \quad E_0^{-}=S^{-(N)}, \quad E_1^{+}=T^{-(N)}, \quad 
E_1^{-}=T^{+(N)}, \quad H_0=-H_1= {\frac 2 N}  S^Z \, .    
\end{equation}
It is shown in Ref. \cite{DFM} 
that the operators $E_j^{\pm}, H_j$ for $j=0,1$,  
satisfy the defining relations  of the algebra $U(L(sl_2))$.

\par
Noticing the automorphism 
\be 
\theta(E_0^{\pm})=E_1^{\pm} \,  , \quad \theta(H_0)=  H_1 
\ee
we may take the  identification in the following:   
\begin{equation}
E_0^{+} = T^{-(N)}, \quad E_0^{-}=T^{+(N)}, \quad E_1^{+} = S^{+(N)}, \quad 
E_1^{-} = S^{-(N)} , \quad H_0=-H_1= - {\frac 2 N}  S^Z \, .    
\label{id2}
\end{equation}

\subsection{Classical analogue of the Drinfeld realization}

Let us review  briefly  the Drinfeld realization of 
the quantum loop algebra \cite{Chari-P1,Chari-P2}. 
The quantum affine algebra $U_q({\hat{sl}}_2)$ is isomorphic to the 
associative algebra over ${\bf C}$ with generators 
$x_{k}^{\pm}$ ($k \in {\bf Z}$), $h_{k}$ 
($k \in {\bf Z} \setminus \{ 0 \}$), 
$K^{\pm 1}$, central element $C^{\pm 1}$, 
and the following relations:   
\bea 
C C^{-1} & = & C^{-1} C = K K^{-1} = K^{-1} K = 1 \non \\ 
 \mbox{[} h_{k}, h_{\ell} \mbox{]} & = & \delta_{k, -\ell} \, 
{\frac 1 k} [2k] {\frac {C^k - C^{-k}} {q-q^{-1}}} \non \\
K h_k & = & h_k K , K x_k^{\pm} K^{-1} = q^{\pm 2} x_k^{\pm} \non \\  
\mbox{[} h_k, x_{\ell}^{\pm} \mbox{]} & = & 
\pm {\frac 1 k} [2k] C^{\mp (k + |k|)/2} 
x_{k+\ell}^{\pm} \non \\
 x_{k+1}^{\pm} x_{k}^{\pm} & -&  q^{\pm 2} x_{\ell}^{\pm} x_{k+1}^{\pm} 
= q^{\pm 2 } x_k^{\pm} x_{\ell+1}^{\pm} - x_{\ell+1}^{\pm} x_{k}^{\pm} \non \\ 
\mbox{ [} x_k^{+}, x_{\ell}^{-} \mbox{]} & = & {\frac 1 {q-q^{-1}} } 
\left( C^{k-\ell} \psi_{k+\ell} - \phi_{k+\ell} \right) \non \\
\sum_{k=0}^{\infty}@\psi_k u^k & = & K 
\exp\left((q-q^{-1})\sum_{k=1}^{\infty} h_k u^k \right)  \non \\
\sum_{k=0}^{\infty}@\phi_{-k} u^{-k} & = & K^{-1} 
\exp\left(-(q-q^{-1})\sum_{k=1}^{\infty} h_{-k} u^{-k} \right)  \non \\
\label{drinfeld}
\eea

%
%

Let us consider the classical analogue of the Drinfeld realization 
\cite{Chari-P2}. Putting $q = \exp \epsilon$, we take the limit of
 $\epsilon$ to zero: $q = 1 + \epsilon + \cdots $.  Hereafter we 
 assume the trivial center: $C^{\pm 1}=1$.  We define the classical 
 analogs of the Drinfeld generators as follows.  
\bea 
K & = & q^{{\bar h}_0} \non \\
h_k & = & {\bar h}_K + O(\epsilon) \qquad (k \in {\bf Z} \setminus \{0 \}) 
\non \\
x_k^{\pm} &= & {\bar x}_k^{\pm} + O(\epsilon) \qquad (k \in {\bf Z})  
\eea
 From the last two relations of the set of defining relations 
 of eqs. (\ref{drinfeld}), 
 we have the following relations 
\bea 
\psi_k & = & 2 \epsilon {\bar h}_k + O( \epsilon^2) \qquad (k \ge 1) \non \\
\phi_{-k} & = & - 2 \epsilon {\bar h}_{-k} + O(\epsilon^2) 
\qquad (k \ge 1) . 
\eea 
Then, from the classical limit of the quantum loop algebra,  
 we have the following relations: 
\bea 
\mbox{[} {\bar h}_k , {\bar h}_{\ell} \mbox{]} & =&  0 \qquad \non \\ 
\mbox{[} {\bar h}_{k}, x_{\ell}^{\pm} \mbox{]} & = & 
\pm 2 x_{k + \ell}^{\pm} \non \\  
{\bar x}_{k+1}^{\pm} {\bar x}_{\ell}^{\pm}  -
{\bar x}_{\ell}^{\pm} {\bar x}_{k+1}^{\pm}
& = & 
{\bar x}_{k}^{\pm} {\bar x}_{\ell+1}^{\pm}  -
{\bar x}_{\ell+1}^{\pm} {\bar x}_{k}^{\pm} \non \\
\mbox{[} {\bar x}_{k}^{+}, {\bar x}_{\ell}^{-}   \mbox{]}
& = & {\bar h}_{k+ \ell} 
\eea
Here $k, \ell \in {\bf Z}$.  
\par 
Let us consider  
the classical limit of the isomorphism of the Drinfeld realization of 
 the quantum loop group  to the quantum affine algebra 
 $U_q({\hat{sl}}_2)$. It is the isomorphism 
 between the $sl_2$ loop algebra and the classical limit of the 
 quantum affine algebra given by the following: 
\bea
& & E_1^{\pm} \mapsto {\bar x}_0^{\pm}  \non \\
& & E_0^{+} \mapsto {\bar x}_1^{-} \qquad  
E_0^{-} \mapsto {\bar x}_{-1}^{+}  \non \\
& & -H_0 =  H_1 \mapsto {\bar h}_0
\eea
Applying the isomorphism to the identification   
(\ref{id2}), we have the correspondence:    
\bea 
{\bar x}_0^{+} = T^{-(N)} \, , \quad  
{\bar x}_0^{-} = T^{+(N)} \, , \quad  
{\bar x}_1^{+} = S^{+(N)}  \, , \quad  
{\bar x}_{-1}^{-} = S^{-(N)} \, , \quad  
{\bar h}_{0} = {\frac 2 N} S^Z 
\eea

\par 
Let us now give  
the definition of a highest weight representation 
for the loop algebra $U(L(sl_2))$.  
They are given by the following:  

A representation of the $sl_2$ loop algebra is highest weight
if it is generated by a vector $\Omega$ which is annihilated 
by the generator ${\bar x}_k^{+}$ for all $k \in {\bf Z}$ and 
such that $\Omega$ is an eigenvector of 
the Cartan generator ${\bar h}_k$ for $k \in {\bf Z}$.   

Thus, a vector $\Omega$ is highest weight if 
it is annihilated by ${\bar x}_k^{+}$ for all integer $k$, 
and it is an eigenvector of the generator ${\bar h}_k$ for 
all integer $k$.

\section{Formulas of algebraic Bethe ansatz}

\subsection{$R$ matrix and $L$ operator}
\par 
Let us summarize  some formulas of the algebraic 
Bethe ansatz \cite{8VABA,TF,Korepin}.
We define the $R$ matrix of the XXZ spin chain by 
\be
R(z-w)  
= \left(
\begin{array}{cccc} 
f(w-z) &   0 & 0 & 0 \\
0 &   g(w-z) & 1 & 0 \\
0 &  1 & g(w-z) & 0 \\
0 &   0 & 0 & f(w-z)
\end{array} 
\right) 
\ee
where $f(z-w)$ and $g(z-w)$ are given by 
\be 
f(z-w)= {\frac {\sinh(z-w-2 \eta)} {\sinh(z-w)}} \quad
g(z-w)= {\frac {\sinh(-2 \eta)} {\sinh(z-w)}} 
\ee
Hereafter we shall write $f(t_1-t_2)$ by $f_{12}$ for short. 

\par 
We now introduce $L$ operators for the XXZ spin chain  
\be 
 L_n(z) 
= \left(
\begin{array}{cc}  
L_n(z)^1_1  &  L_n(z)^1_2 \\
 L_n(z)^2_1  & L_n(z)^2_2  
\end{array} 
\right)  
=  \left(
\begin{array}{cc}  
\sinh \left( z I_n + \eta \sigma_n^z \right) & \sinh 2 \eta \, \sigma_n^{-} \\
\sinh 2 \eta \, \sigma_n^{+}  & \sinh \left( z I_n - \eta \sigma_n^z \right) 
\end{array} 
\right)  
\ee
Here $I_n$ and $\sigma_n^a$ ($n=1, \ldots, L$) are 
acting on the $n$ th vector space $V_n$. 
The $L$ operator is an operator-valued matrix 
which acts on the auxiliary vector space 
$V_0$.  The symbols $\sigma^{\pm}$ denote 
$\sigma^{+}= E_{12}$ and $\sigma^{-} = E_{21}$,    
and $\sigma^x, \sigma^y, \sigma^z$ are the Pauli matrices  

\par 
In terms of the $R$ matrix and $L$ operators, 
the Yang-Baxter equation is expressed as 
\be 
R(z-t) \left( L_n(z) \otimes L_n(t) \right) 
 = \left( L_n(t) \otimes L_n(z) \right) R(z-t) 
\label{RLL} 
\ee
We define the monodromy matrix $T$ by the product: 
$ T(z) = L_L(z) \cdots L_2(z) L_1(z)$,  
and the transfer matrix $\tau_{6V}(z)$  
by the trace: $\tau_{6V}(z) = {\rm Tr} \, T(z)$.
The matrix elements of $T(z)$   
\be 
T(z) = 
\left( \begin{array}{cc}
A(z) & B(z) \\
C(z) & D(z) 
\end{array} 
\right)
\ee
satisfy the commutation relations derived from 
the Yang-Baxter equations such as 
$B_1 B_2 = B_2 B_1$ and 
\be 
A_1 B_2 = f_{12} B_2 A_1 - g_{12} B_1 A_2 \, , \quad 
D_1 B_2 = f_{21} B_2 D_1 - g_{21} B_1 D_2 
\label{CR}
\ee
Here we recall $B_1$ denotes $B(t_1)$.  
Furthermore we can show \cite{Korepin}
\bea 
C_0 B_1 \cdots B_n & = & B_1 B_2 \cdots B_n C_0 \non \\
& + & \sum_{j=1}^{n} B_1 \cdots B_{j-1} B_{j+1} \cdots B_n \, 
 g_{0j} \left\{ A_0 D_j \prod_{k \ne j} f_{0k} f_{kj} 
- A_j D_0 \prod_{k \ne j} f_{k0} f_{jk} 
\right\} \non \\
& - & \sum_{1 \le j < k \le n} 
B_0 B_1 \cdots B_{j-1} B_{j+1} \cdots B_{k-1}B_{k+1} \cdots B_n  \non \\ 
& & \quad \times g_{0j} g_{0k} \{A_j D_k f_{jk} 
\prod_{\ell \ne j, k} f_{j \ell}f_{\ell k}  
+ A_k D_j f_{kj} \prod_{\ell \ne j,k } f_{k \ell} f_{\ell j} \} 
\label{CBBB}
\eea

Let us denote by $|0 \ra$ the vector where all the spins are up. 
Then we can show 
\be 
A(z) | 0 \ra = a(z) | 0 \ra \, , \quad 
D(z) | 0 \ra = d(z) | 0 \ra 
\ee
where $a(z)= \sinh^L(z+ \eta)$ and $d(z)=\sinh^L(z-\eta)$ . 

\par 
Making use of the commutation relations (\ref{CR}) 
one can show that the vector $B_1 B_2 \cdots B_R | 0 \ra$ is  
an eigenvector of the XXZ spin Hamiltonian if the rapidities 
$t_1, t_2, \ldots, t_R$ satisfy the Bethe ansatz equations  
\be 
{\frac {a_j} {d_j}} = \prod_{k \ne j} 
\left( {\frac {f_{kj} } {f_{jk}}} \right) 
\quad {\rm for} \quad j= 1, \cdots R \, .  
\label{BAE}
\ee

\subsection{Formulas with  infinite rapidities}

Let us normalize the operators $A(z)$ 's as follows 
\bea 
{\hat A}(z) & = & A(z)/n(z) \qquad {\hat B}(z) = B(z)/(g(z) n(z)) \non \\  
{\hat D}(z) & = & D(z)/n(z) \qquad {\hat C}(z) = B(z)/(g(z) n(z))  \, . 
\eea
Here the normalization factor $n(z)$ is given by $n(z)= \sinh^L z$. 
Then, we can show the following: 
\bea 
{\hat A}(\pm \infty) & = & q^{\pm S^Z} \qquad {\hat B}(\infty) 
= - T^{-} \quad  {\hat B}(-\infty) = - S^{-} 
 \non \\    
{\hat D}(\pm \infty) & = & q^{\mp S^Z} \, \quad  {\hat C}(\infty) 
= - S^{+} \quad  {\hat C}(-\infty) = - T^{+} 
\eea
Here, we recall that $S^{\pm}$ and $S^Z$ are defined on the $L$th tensor 
product of spin 1/2 representations.  
Hereafter, we shall write ${\hat C}(\infty)$'s simply as  
 ${\hat C}_{\infty}$'s.

\par 
We now derive  quite a useful formula. Sending $z_0$ to infinity 
in eq. (\ref{CBBB}),  we have the following:  
\bea 
{\hat C}_{\infty} \, B_1 \cdots B_M \ 0 \ra & = & 
\sum_{j=1}^{M} B_1 \cdots B_{j-1} B_{j+1} \cdots B_M | 0 \ra \non \\
& \times & e^{t_j} \left( q^{(L/2) - (M-1)} d_j 
\prod_{k \ne j} f_{kj} - q^{-(L/2)+ (M-1)} 
a_j \prod_{k \ne j} f_{jk}  \right)
\label{formulaCB} 
\eea
where $t_1 , \ldots, t_M$ are given arbitrary. 
Applying the formula (\ref{formulaCB}) $N$ times, we have   
an important formula in the  following: 
\bea 
& &  \left( {\hat C}_{\infty} \right)^{N} B_1 \cdots B_M | 0 \ra \non \\ 
& = & \sum_{S_N \subseteq \Sigma_M} \sum_{P \in {\cal S}_N} 
\left( \prod_{j \in \Sigma_M \setminus S_N}  B_{j} \right) \,  
|0 \ra \, \exp( \sum_{j \in S_N} t_j ) \non \\
& & \prod_{n=1}^{N} \left( q^{(L/2) - (M-n) } d_{j_{P n}} 
\prod_{k \ne j_{P1}, \cdots, j_{Pn}} f_{k j_{Pn}} 
- q^{-(L/2) + (M-n) } a_{j_{P n}} 
\prod_{k \ne j_{P1}, \cdots, j_{Pn}} f_{j_{Pn} k} \right)
\label{CCC}
\eea
Here the set $\Sigma_M$ is given by $\Sigma_M= \{ 1, 2, \cdots, M \}$ and 
$S_N$ denote the subset of $\Sigma_M$ with $N$ elements.  
The symbol ${\cal S}_N$ denotes the symmetry group of $N$ elements. 
The symbol $P \in {\cal S}_N$ means that $P$ 
is a permutation of $N$ letters.  
Here we recall that $t_1 , \ldots, t_M$ are given arbitrary.

\section{Proof of the annihilation 
of regular XXZ Bethe vectors by the Drinfeld generators} 

\subsection{The $N$th power of $C$ operators acting on regular Bethe states }

\par 
Let us now discuss the proof for the annihilation property: 
when $t_1, \ldots, t_R$ are finite and 
they satisfy the Bethe ansatz equations (\ref{BAE}), 
then we have 
\be 
S^{+(N) } \, B_1 \cdots B_R | 0 \ra = 0  
\ee  
We may assume $R > N$, otherwise it is trivial.  
Making use of the Bethe ansatz equations (\ref{BAE}) we have 
the following  formula: 
\bea 
& & \prod_{n=1}^{N} \left( q^{(L/2) - (R-n) } d_{j_{P n}} 
\prod_{k \ne j_{P1}, \cdots, j_{Pn}} f_{k j_{Pn}} 
- q^{-(L/2) + (R-n) } a_{j_{P n}} 
\prod_{k \ne j_{P1}, \cdots, j_{Pn}} f_{j_{Pn} k} \right) \non \\
& = & \left( \prod_{\ell=1}^{N} a_{j_{P \ell}} \right) 
\prod_{n=1}^{N} \left( \prod_{k \notin S_N} f_{j_{Pn} k} \right)  \,
\prod_{n=1}^{N} \left( q^{(L/2) - (R-n) }  
\prod_{k \ne j_{P1}, \cdots, j_{Pn}} f_{j_{Pn} j_{P \ell}} 
- q^{-(L/2) + (R-n) } 
\prod_{\ell=1}^{n-1} f_{j_{P \ell} j_{Pn}} \right)
\eea

By induction on $N$, we can show the following  formula: 
\be 
 \sum_{P \in {\cal S}_N} \prod_{n=1}^{N} \left( x_n \prod_{\ell=1}^{n-1} 
f_{j_{Pn} j_{P \ell}} - y_n 
\prod_{\ell=1}^{n-1} f_{j_{P\ell} j_{Pn}} \right)     
 =  \prod_{n=1}^{N} (x_n -y_n) \times \sum_{P \in {\cal S}_N } 
\prod_{n=1}^{N} \prod_{\ell=1}^{n-1} f_{j_{P\ell} j_{Pn}}
\label{product} 
\ee

Applying the formula (\ref{product}) we obtain  
\bea 
& &  \left( {\hat C}_{\infty} \right)^{N} \, B_1 \cdots B_R | 0 \ra \non \\ 
& = & \sum_{S_N \subseteq \Sigma_R} 
\left( \prod_{\ell \in \Sigma_R \setminus S_N}  B_{\ell} \right) \,  
|0 \ra \, \exp( \sum_{j \in S_N} t_j ) \, 
\left( \prod_{j \in S_N} a_j \right) \prod_{j \in S_N} \left( 
\prod_{k \notin S_N} f_{jk} \right)   \non \\
& & 
\prod_{n=1}^{N} \left( q^{(L/2) - (R-n) } - q^{-(L/2) +R -n} \right) \, 
\sum_{P \in {\cal S}_N} 
\prod_{n=1}^{N}  
\prod_{\ell=1}^{n-1} f_{j_{P \ell} j_{Pn}}
\label{CN}
\eea
Noting the relation 
\be 
{\hat C}(\infty)^N = (-1)^N \, \left( S^{+} \right)^N \, , 
\ee
we divide the both hand sides of (\ref{CN}) by the $q$ factorial $[N] !$,    
and we have for the case of generic $q$ 
\bea 
& &  S^{+ (N)} \, B_1 B_2 \cdots B_R | 0 \ra \non \\ 
& = & (-1)^N \sum_{S_N \subseteq \Sigma_R} \sum_{P \in {\cal S}_N} 
\left( \prod_{\ell \in \Sigma_R \setminus S_N}  B_{\ell} \right) \,  
|0 \ra \, \exp( \sum_{j \in S_N} t_j ) \, 
\left( \prod_{j \in S_N} a_j \right) \prod_{j \in S_N} \left( 
\prod_{k \notin S_N} f_{jk} \right)   \non \\
& & 
\times 
\left[ 
\begin{array}{c}
{\frac L 2} - R + N \\ 
N 
\end{array}
\right]_q
\, (q-q^{-1})^N  
\, \left( \sum_{P \in {\cal S}_N} 
\prod_{n=1}^{N}  
\prod_{\ell=1}^{n-1} f_{j_{P \ell} j_{Pn}} \right)
\label{SB}
\eea
Here, we have defined the q-binomial by 
\be 
\left[
\begin{array}{c} 
m \\
n
\end{array} 
 \right]_q 
 = {\frac {[m] !} {[m-n]! \, [n]!} } \quad , \, 
 \mbox{ for } \quad m \ge n \, . 
\ee

\subsection{The case of roots of unity}
\par 
Let us now consider the case when $q$ is a root of unity. 
When $q$ is a $2N$th root of unity or an $N$th root of unity with $N$ odd, 
we can show the following equality  
\be 
 \sum_{P \in {\cal S}_N} 
\prod_{n=1}^{N}  
\prod_{\ell=1}^{n-1} f_{j_{P \ell} j_{Pn}}
= 0 
\label{symsum}
\ee
Let us define a function $F(z_1, \cdots, z_N)$ by 
\be 
F(z_1, \cdots, z_N) = \sum_{P \in {\cal S}_N} 
\prod_{n=1}^{N}  
\prod_{\ell=1}^{n-1} f_{{P \ell},{Pn}}
\ee
Then  we can show that there is no pole for $F(z_1, \cdots, z_N)$ 
with respect to any variable $z_j$,  
and also that $F(z_1, \cdots, z_N)$  vanishes when one of the variables 
is sent to infinity: $z_j = \infty$ for some $j$. Therefore we have  
the equality (\ref{symsum}). It follows from 
 (\ref{SB}) and (\ref{symsum}) that the action of the operator $S^{+(N)}$ 
on the regular Bethe ansatz state 
$ B_1 \cdots B_R |0 \ra $ is given by zero. 

\par 
By a similar method with $S^{+(N)}$, we can show 
that the operator $T^{+(N)}$ makes the regular Bethe ansatz eigenstate 
vanishes. Thus, we have obtained 
\be 
S^{+(N)} \, B_1 \cdots B_M |0 \ra = 0 \, ,  \quad 
T^{+(N)} \, B_1 \cdots B_M |0 \ra = 0. 
\label{vanish}
\ee

\section{Eigenvalues of the Cartan generators  
on regular Bethe states}

We  discuss a method by which we can calculate the 
eigenvalue of the generator ${\bar h}_k$ 
on a regular Bethe state. For an illustration, 
we consider the case of $N=2$. 

Let us calculate the eigenvalue of 
the operator $S^{+(2)} T^{-(2)}$ on a regular Bethe state with 
$R$ down-spins: $| R \ra = B_1 \cdots B_R |0 \ra$.  
Here we recall that $R$ denotes the number of regular
 rapidities.  We set $M=R+2$. 
We assume that $t_1, \ldots, t_R$ satisfy the Bethe ansatz equations, 
while $z_{R+1}$ and $z_{R+2}$ are sent to infinity, later. 
We now consider the formula (\ref{formula}) for the case of 
$N=2$ in the following: 
\bea 
\left( 
{\hat C}_{\infty}
\right)^2 
B_1 B_2 \cdots B_M | 0 \ra & = & \sum_{S_2 \subset \Sigma_M} 
 \sum_{P \in {\cal{S}}_2} \prod_{\ell \in \Sigma_M \setminus S_2} 
 B_{\ell} |0 \ra \, \exp(\sum_{j \in S_2}) \non \\ 
& \times  & \prod_{n=1}^{2} \left(a_{j_{Pn}} q^{-L/2} 
\prod_{k \ne j_{P1}, \ldots, j_{Pn}} (q f_{j_{Pn} \, k}) - 
d_{j_{Pn}} q^{L/2} 
\prod_{k \ne j_{P1}, \ldots, j_{Pn}}(q^{-1} f_{k \, j_{Pn}})  \right)  
\label{N=2}
\eea
Here we recall $T^{-} = - {\hat B}(\infty)$ and $S^{+} 
= - {\hat C}(\infty)$. In eq. (\ref{N=2}) we have  four cases 
for the set $S_2$: $S_2 = \{ R+1, R+2 \}$, $\{j, R+2 \}$, 
$\{j, R+1 \}$ and $\{ j_1, j_2\}$. Here, $j, j_1, j_2 \in \{ 
1, 2, \ldots, R \} $. For an illustration,  we highlight the term 
with $S_2 = \{ R+1, R+2 \}$ in the following:   
\bea 
& & \left( 
{\hat C}_{\infty}
\right)^2 \prod_{\ell \in \Sigma_M} 
B_{\ell} |0 \ra \non \\
& = & \prod_{\ell=1}^{R} 
B_{\ell} |0 \ra \, e^{z_{R+1} + z_{R+2}} \,  \sum_{P \in {\cal S}_2} 
\prod_{n=1}^{2} \left(a_{j_{Pn}} q^{-L/2} 
\prod_{k \ne j_{P1}, \ldots, j_{Pn}} (q f_{j_{Pn} \, k}) - 
d_{j_{Pn}} q^{L/2} 
\prod_{k \ne j_{P1}, \ldots, j_{Pn}}(q^{-1} f_{k \, j_{Pn}})  \right)  
\non \\
& + & \cdots . 
 \eea

After sending $z_{R+1}$ to infinity,  
we take the limit of sending $z_{R+2}$ to infinity. 
 Then, we have the following:  
\bea 
& & \lim_{z_{R+2} \rightarrow \infty} 
\left( \lim_{z_{R+1} \rightarrow \infty} \left( {\hat C}_{\infty} \right)^2 
B_1 \cdots B_R {\hat B}_{R+1} {\hat B}_{R+2} | 0 \ra \right) \non \\
& = & \prod_{\ell=1}^{R} B_{\ell} | 0 \ra \, [2]^2 \left\{ 
 \left(_L C_2 +[3] \sum_{k=1}^{R} e^{4t_k} \right)   
- [2] L \sum_{k=1}^{R} e^{2 t_k} + [2]^2 \sum_{j < k} e^{2(t_j + t_K)}  
 \right\} \non \\ 
& + & \sum_{j=1}^{R} 
\left[ {\hat B}_{\infty} \prod_{\ell \ne j} B_{\ell} | 0 \ra 
\left( - (q^2 + q^{-2} ) \sum_{k \ne j} e^{2t_k} 
+ (L-1) - e^{2 t_j} \right) - {\frac {\delta {\hat B}_{\infty}} 
{\delta \epsilon_{R+2}} } \prod_{\ell \ne j} B_{\ell} | 0 \ra 
   \right] \non \\
& & \times [{\frac L 2}-R-2] [2]^2 (q-q^{-1}) e^{t_j} a_j 
\prod_{k\ne j} f_{jk}    \non \\ 
& + & 
  \sum_{1 \le j_1 < j_2 \le R} 
  \left( {\hat B}_{\infty} \right)^2  
   \prod_{\ell = 1; \ell \ne j_1, j_2}^{R}    
   B_{\ell} | 0 \ra \, e^{t_{j_1} + t_{j_2}} (q-q^{-1})^2 \non \\
& &   \times [{\frac L 2} - R - 2]  [{\frac L 2} - R - 3] [2] \, 
a_{j_1} a_{j_2} \prod_{k \ne j_1, j_2} f_{j_1\, k} f_{j_2 \, k}@ 
\label{CCBB}
 \eea  
Here, $\epsilon_{R+2} = \exp(-2 z_{R+2})$.   
We note that $q$ is generic, so far. 

\par 
We multiply the normalization factor $[2]^2$ with (\ref{CCBB}). 
Then we take the limit of sending  $q$ to a root of unity. 
If $L/2 - R \equiv 0$ (mod 2), then all the unwanted terms vanish.  
We have 
\be 
S^{+ (2)} T^{-(2)} \, B_1 \cdots B_R | 0 \ra = 
\left(_L C_2 +[3] \sum_{k=1}^{R} e^{4t_k} \right)    
B_1 \cdots B_R | 0 \ra 
\ee

\section{Classical analogue of the Drinfeld polynomial}

\subsection{Definition}

Let us denote 
the eigenvalue of the generator ${\bar h}_k$ on $\Omega$  as 
\be 
{\bar h}_k \Omega = {\bar d}_k^{+} \Omega   
\qquad 
- {\bar h}_{-k} \Omega = {\bar d}_{-k}^{-} \Omega   
\ee
for $k \ge 1 $ and $k  \in {\bf Z}$.

The classical analogue of the Drinfeld polynomial $P(u)$ satisfies   
\be 
\sum_{k=1}^{\infty} {\bar d}_k u^k = -u {\frac {P^{'}(u)} {P(u)}} 
 \, , \quad  
\sum_{k=1}^{\infty} {\bar d}_{-k} u^{-k} = deg P 
-u {\frac {P^{'}(u)} {P(u)}}  \, .  
\ee

\subsection{Some useful formulas}

Recall that ${\bar x}_k^{\pm}$ and ${\bar h}_k$ denote 
the classical analogues 
of the Drinfeld generators ${x}_k^{\pm}$ and ${h}_k$, respectively.   
Then, we can show 
\be 
{\bar x}_k^{+} = {\frac 1 {2^k}} 
\left( ad_{{\bar h}_1} \right)^{k} {\bar x}_{0}^{+} \, , \quad 
{\bar x}_{-k}^{+} = {\frac 1 {2^k}} 
\left( ad_{{\bar h}_{-1}} \right)^{k} {\bar x}_{0}^{+} 
\label{xx}
\ee
for $k > 0$. Here ${\bar h}_1$ is given by 
\be 
{\bar h}_1 = [{\bar x}_0^{+}, {\bar x}_1^{-}]  \, , 
\quad  {\bar h}_{-1} = (-1) [{\bar x}_{-1}^{+}, {\bar x}_0^{+}]  \, , 
\ee
From the second identification (\ref{id2}) we have 
\be 
{\bar h}_1 = [S^{+(N)}, T^{-(N)}] \, , \quad
{\bar h}_{-1} = [S^{-(N)}, T^{+(N)}] \, , 
\ee
In the second identification (\ref{id2}), 
we have 
\be 
{\bar x}_0^{+} = S^{+(N)} 
\ee
Thus, we can express the generators ${\bar x}_{k}^{+}$ 
in terms of $S^{\pm (N)}$ and $T^{\pm (N)}$ for any integer $k$. 

\par 
 The equation (\ref{xx}) can also be formulated as 
 \be 
 \mbox{[} {\bar h}_1, {\bar x}_k  \mbox{]} = 2 {\bar x}_{k+1}^{+} 
\label{rec}
 \ee
 Making use of eqs. (\ref{rec})  and (\ref{vanish}), 
 we can recursively  show that 
 the generator ${\bar x}_k^{+}$ 
 vanishes on  regular XXZ Bethe ansatz eigenvectors at roots of unity.  
 Here, we have assumed that 
 the action of ${\bar h}_1$ on a regular Bethe ansatz state 
 can be calculated  with the method of \S 5. For $N=2$, 
 it is finished as shown in \S 5. Precisely speaking, however, 
 it is a conjecture for $N > 2$.   
 
\par 
In order to calculate the evaluation parameters of the Drinfeld polynomial, 
we can use the following: 
\be 
{\bar h}_{k+1} = {\frac 1 2} 
[ [ {\bar h}_k, {\bar x}_0^{+}], {\bar x}_1^{-} ]    
\ee

\subsection{An example of the classical analogue of the Drinfeld polynomial}

For $L=6$ and $N=3$, 
the Drinfeld polynomial of the degenerate eigenspace for $| 0 \ra$ 
is given by 
\be 
P(u) = (1-a_1 u) (1-a_2 u )
\ee
 where $a_1$ and $a_2$ are given by  $10\pm 3 {\sqrt{11}}$. 
 The roots are distinct and the degree of $P$ is two,
  so that the dimension is given by $2^2=4$.  

\section{Discussion}

Through the classical analogue of the Drinfeld polynomial, 
we can calculate the dimension of the multiplets of the $sl_2$ loop algebra. 
\par 
The dimensions of the $sl_2$ multiplets should be consistent with 
that of the XYZ model: $2^{(L-2M)/N}$ \cite{Missing,CSOS}. 
In fact, we can show that when
$L = 2M$ and the rapidities $t_1 \ldots, t_M$ are 
finite-valued solutions to the Bethe ansatz equations we have  
\be 
S^{ \pm (N)} B_1 \cdots B_M | 0 \ra = T^{ \pm (N)} B_1 \cdots B_M | 0 \ra = 0  
\ee 

\par 
More details of computation of the 
classical analogues of the Drinfeld polynomials should be 
reported elsewhere. 

\par 
%

{\vskip 1.2cm}
\par \noindent 
{\bf Acknowledgements}


This work is partially supported by the Grant-in-Aid (No. 14702012).

\end{document}